# Design of Black Phosphorus 2D Nanomechanical Resonators by Exploiting the Intrinsic Mechanical Anisotropy

Zenghui Wang, Philip X.-L. Feng[*]

*Department of Electrical Engineering & Computer Science, Case School of Engineering,*

*Case Western Reserve University, 10900 Euclid Avenue, Cleveland, OH 44106, USA*

## Abstract

**Black phosphorus (P), a layered material that can be isolated down to individual 2D crystalline sheets, exhibits highly anisotropic mechanical properties due to its corrugated crystal structure in each atomic layer, which are intriguing for 2D nanomechanical devices. Here we lay the framework for describing the mechanical resonant responses in free-standing black P structures, by using a combination of analytical modeling and numerical simulation. We find that thicker devices (>100nm) operating in the elastic plate regime exhibit pronounced signatures of mechanical anisotropy, and can lead to new multimode resonant characteristics in terms of mode sequences, shapes, and orientational preferences that are unavailable in nanomechanical resonators made of isotropic materials. In addition, through investigating devices with different geometries, we identify the resonant responses' dependence on crystal orientation in asymmetric devices, and evaluate the effects from the degree of anisotropy. The results suggest a pathway towards harnessing the mechanical anisotropy in black P for building novel 2D nanomechanical devices and resonant transducers with engineerable multimode functions.**

---

[*]Corresponding Author. Email: philip.feng@case.edu

Ultrathin crystalline black phosphorus (P), a material with atomic P layers stacked via van der Waals force, has recently joined the family of two-dimensional (2D) nanomaterials, bringing with it a number of unique properties unavailable in other 2D crystals, such as the corrugated/puckered structures in each atomic layer, and the thickness-tunable bandgap accessing the infrared regime [1,2,3,4,5,6]. Among these properties, a compelling feature is the strong anisotropy in the material's physical properties [7,8,9], which stems from its special crystalline structure: inside a single layer, each P atom is covalently bonded with three adjacent P atoms to form a corrugated plane of honeycomb structure (Fig. 1a), differentiating the two orthogonal in-plane directions—parallel and perpendicular to the atomic corrugations (*i.e.*, $x$ and $y$ directions in Fig. 1a). The strong anisotropy in the material's electronic, optical, and optoelectronic properties has led to the predictions and observations of a variety of new phenomena in black P nanostructures [3,4,10,11,12,13,14,15,16].

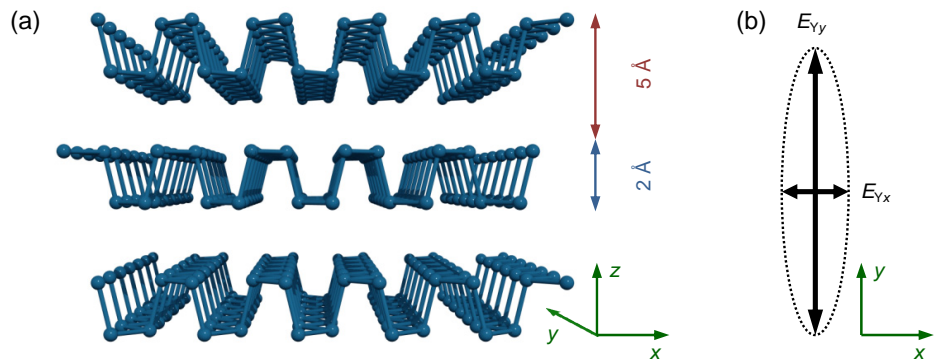

**Figure 1. 2D black phosphorus material structure and mechanical anisotropy.** (a) Schematic of black P crystal. (b) Anisotropic in-plane elastic moduli of black P.

In the mechanical domain, the strong anisotropy is manifested in the highly-directional elastic modulus—which can vary by a factor of ~5 in the two in-plane directions [7,8] (see Fig. 1b, $E_{Y,x} \approx 37\text{-}41.3\,\text{GPa}$ and $E_{Y,y} \approx 106.4\text{–}159\,\text{GPa}$)—may lead to crystal-orientation-dependent behavior in nanomechanical devices based black P. In particular, the strong mechanical anisotropy found in black P is unavailable in other materials used for MEMS/NEMS devices: for example, Si, the most widely used material for micro- and nanomechanical structures whose mechanical anisotropy is well-studied [17], exhibits at most 50% difference in Young's modulus along different crystal orientations [18], one order of magnitude smaller than that found in black P. Further, as Si-based devices are thinned down to 100nm or less, its mechanical anisotropy may not be preserved as the crystal properties give way to surface effects and lattice defects [19,20,21], which can be introduced during the subtractive nanofabrication processes. In contrast, the layered structure of black P preserves the anisotropy all the way down to a monolayer [7], as the naturally-cleaved devices surfaces often remain atomically perfect, with each layer being intrinsically anisotropic. Therefore, black P offers truly unique properties for exploring nanomechanical devices based on materials with strong mechanical anisotropy. In this work, we present a theoretical framework to investigate the frequency scaling of black P nanomechanical resonators, elucidate its contrast with devices based on other 2D materials, and explore the material's mechanical anisotropy for device designs.

As demonstrated in recent experiments [22], robust resonant responses exist in black P nanomechanical resonators with a variety of device geometries. Previous studies show that nonideal and asymmetric device structures based on 2D materials present new opportunities for engineering multimode resonances [23] toward sensing and signal processing. Here we focus on regular device geometries, *i.e.*, fully clamped drumhead resonators, and their linear resonant responses, to keep the nanomechanical theory concise, while highlighting the new effects created by the anisotropic mechanical properties. The resonance frequency $f_{\text{res}}$ of a fully clamped resonator generally has two components: (i) a membrane part, which exemplifies the effect from the tension; and (ii) a tensionless plate part, which reflects the contribution from the bending (flexural) rigidity. Depending on the tension level and device geometry, a 2D resonator may operate in the membrane limit, the plate limit, or the transition regime in between [24]. Therefore, to capture the entire operating range when examining frequency scaling, it is necessary to use the model of tensioned plates.

We start with the simplest case of circular drumhead devices. The high symmetry in this geometry allows black P resonators with different crystal orientations to become mechanically equivalent in terms of frequency scaling, and makes it possible to make fair comparison across resonators based on various 2D materials. For a tensioned plate resonator, its resonance frequency $f_{res}$ can be expressed as [25,26]:

$$f_{res} = \frac{\left(k_2^{mn}a\right)}{2\pi}\sqrt{\frac{D}{\rho a^4}\left[\frac{\gamma a^2}{D}+\left(k_2^{mn}a\right)^2\right]} = \sqrt{f_{membrane}^2 + f_{plate}^2} \ , \tag{1}$$

where $a$ is the radius of the plate ($a=d/2$ with $d$ being the diameter), $\rho$ is the areal (2D) mass density ($kg/m^2$) of the material, $\gamma$ is the tension (force per unit length, in N/m, as in surface tension) inside the plate. $D=E_Y t^3/[12(1-\nu^2)]$ is the bending rigidity, with $E_Y$ being the Young's modulus, $t$ the thickness of the disk, and $\nu$ the Poisson's ratio. $k_2^{mn}$ is a mode-dependent parameter that is usually solved numerically (details can be found in Ref. 26). As shown in Eq. 1, the expression can be grouped into two simple terms, $f_{membrane}$ and $f_{plate}$, which stand for its frequencies in the membrane and tensionless plate limits, respectively. Eq. 1 shows that $f_{res}$ approaches either limits when the contribution from that limit significantly exceeds that from the other: in the limit of $\gamma a^2/D \rightarrow \infty$ (high tension and/or low bending rigidity), $f_{res} \approx f_{membrane}$; in the other limit $\gamma a^2/D \rightarrow 0$, $f_{res} \approx f_{plate}$. Such elastic responses, especially the transition between the different elastic regimes, have been well observed in nanomechanical resonators based on layered 2D materials [24].

The resonance frequency in the membrane limit can be calculated analytically. For a circular drumhead resonator [27],

$$f_{membrane} = \frac{\alpha_{n,s}}{\pi d} \sqrt{\frac{\gamma}{\rho}}, \qquad (2)$$

where $\alpha_{n,s} = k_2^{mn} a$ is also a mode-dependent parameter ($n$, $s$ refer to the numbers of nodal diameters and nodal circles, respectively; *e.g.*, $\alpha_{0,0} = 2.404$).

The calculation for the tensionless plate limit, in contrast, is much more complicated. For a fully-clamped circular plate resonator with orthotropic elastic modulus, the computation of resonance frequency involves solving a number of transcendental equations including multiple Bessel functions, and for each resonance mode and each parameter set (disk dimension, material density, elastic moduli, Poisson's ratio) the solution must be individually computed, with only a handful of numerical results exist in the literature for a very limited set of parameters [28]. Therefore, in order to investigate the frequency scaling of black P resonators over a large range of dimensions, we use finite element modeling (FEM, using COMSOL) to compute $f_{plate}$ for black P circular drumhead resonators. We then combine the results from the two limits using Eq. 1, and plot it as a function of black P thickness $t$, circular device diameter $d$, and device initial tension $\gamma$.

Figure 2 shows the frequency scaling of the fundamental mode resonance in black P circular drumhead resonators. The $f_{res}$ dependence on device thickness and diameter (with device tension fixed at $\gamma = 1$N/m, a typical value as found in actual devices [22]) is shown in the 3D surface plot (Fig. 2a), with the three different color surfaces representing $f_{\text{membrane}}$ (cyan), $f_{\text{plate}}$ (orange), and $f_{\text{res}}$ (pink). From the 3D plot it can be clearly seen that the devices with larger diameter/smaller thickness behave more membrane-like, while those with opposite attributes tend to operate in the tensionless plate (disk) limit. To quantify this elastic transition in black P resonators, we project the bottom surface (*x-y* plane, colored in grey) of Fig. 2a into a 2D surface (Fig. 2b), and color each pixel according to the relative magnitude of $f_{\text{membrane}}$ and $f_{\text{plate}}$: if one contribution is one order of magnitude greater than the other, the pixel on the parameter surface is colored for that limit (and vice versa), with the rest colored as the transition region. Figure 2b thus clearly shows the transition from one elastic limit to the other and outlines the boundary of the transition region. We find that the boundaries of the transition region closely follow the slope of the curve family $d \propto t^{3/2}$, which is consistent with the theoretical expectation that $\gamma a^2/D$=constant (note that $D \propto t^3$) for the ratio to be not in either limits ($\infty$ or 0), showing good agreement with analytical calculation and FEM simulation.

In the membrane limit, the material's elastic moduli have little effect on its resonance frequency, as $f_{\text{res}}$ is dominated by the built-in tension. Consequently, in terms of frequency scaling, the mechanical anisotropy is only manifested when the device operates in the plate (disk) limit, in which the elastic properties have a strong effect on $f_{\text{res}}$. Accordingly, Fig. 2b also illustrates in this 2D parameter space the domain in which mechanical anisotropy can be clearly observed through the device's resonant responses (the "plate (disk)" area). It clearly shows that, in the practical diameter range (*d*=1-10μm, relevant to current nanofabrication techniques and most device applications), single- and few-layer black P nanomechanical resonators are in the membrane limit, thus making it very challenging to observe any effect on resonance frequency scaling from the material's mechanical anisotropy. In contrast, with typical lateral device sizes and tension levels, multilayer black P resonators (with $t>100$nm) operate in the plate (disk) limit, and are thus the best candidate for utilizing device resonant behavior to reveal and explore the mechanical anisotropy in this corrugated atomic layer 2D material. We note that in in realistic devices [22,24, and references therein], the tension level is found to be on the order of 0.1N/m. Nevertheless, in the extreme theoretical limit of absolute zero tension, even monolayer devices could in principle operate in the plate regime. In such cases, the out-of-plane flexural motion in the tensionless limit is determined by the bending rigidity of the material. Surprisingly, the elastic bending modulus of monolayer black P is expected to be largely isotropic despite the corrugated crystal structure [29]. In contrast, in thicker devices the flexural motion is manifested as expansion and contraction of materials on either side of the neutral plane, and thus the bending rigidity is determined by the Young's modulus of black P, which is highly anisotropic.

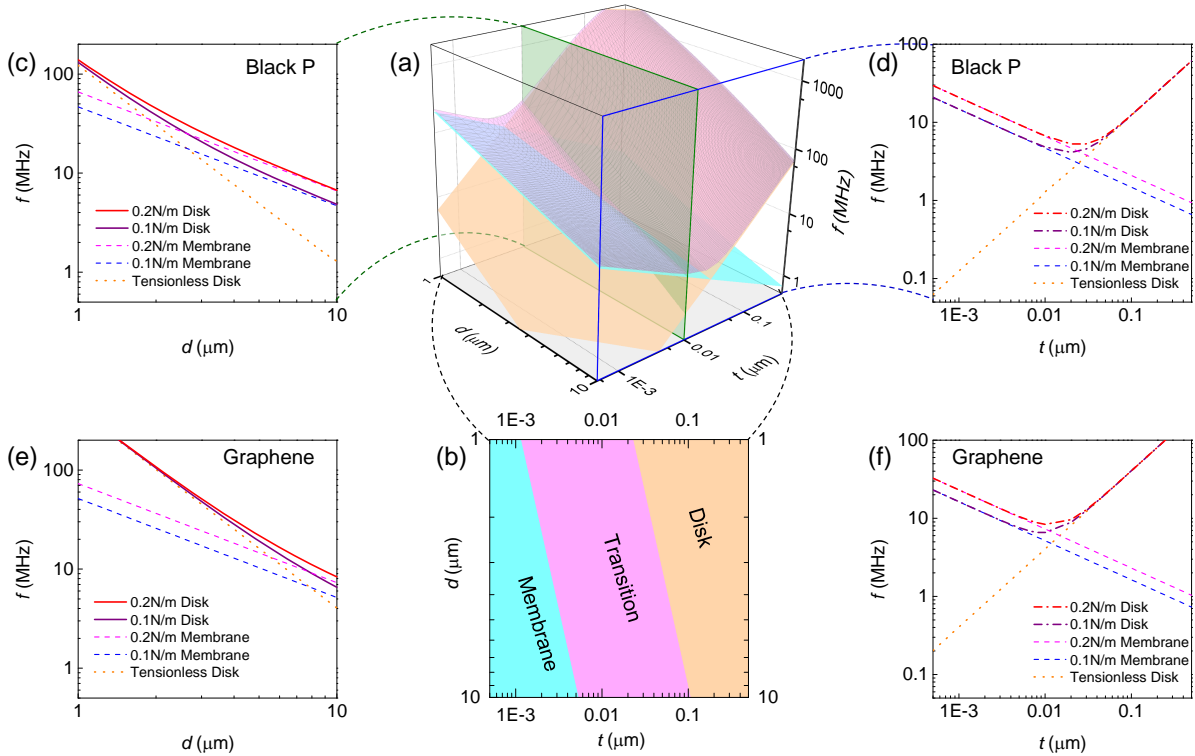

**Figure 2. Frequency scaling of circular black P nanomechanical resonators.** (a) $f_{res}$ (magenta), $f_{membrane}$ (cyan), and $f_{plate}$ (orange) as functions of diameter $d$ and device thickness $t$ for circular drumhead black P nanomechanical resonators with 2D tension $\gamma$=0.1N/m. (b) The $d$-$t$ plane (bottom surface in (a)) showing different elastic regimes determined by the relative magnitude of $f_{\text{membrane}}$ and $f_{\text{plate}}$. (c) Frequency as a function of diameter $d$ for $t$=10nm devices (green vertical surface in (a)) with $\gamma$=0.1 and 0.2N/m. (d) Frequency as a function of thickness $t$ for $d$=10μm devices (blue outlined surface in (a)) with $\gamma$=0.1 and 0.2N/m. (e) The equivalent of (c) for graphene. (f) The equivalent of (d) for graphene.

To evaluate the effect of device tension on the frequency scaling of black P circular drumhead resonators, we examine different 2D cross-sections from the 3D plot (Fig. 2a): the $t$=10nm vertical plane (green surface with outlines) is extracted to generate the $f_{\text{res}}$ *versus d* 2D plot (Fig. 2c), and the $d$=10μm plane (with blue outlines) is used to produce the $f_{\text{res}}$ *versus t* plot (Fig. 2d). In both plots we show $f_{\text{membrane}}$, $f_{\text{disk}}$, and $f_{\text{res}}$, with $\gamma$ varying between 1N/m and 2N/m. Fig. 2c & d show that devices with larger diameters and/or smaller thicknesses are more likely in the membrane limits. It can also be seen that $f_{\text{disk}}$ does not depend on device tension, while $f_{\text{membrane}}$ increases with $\gamma$. The increase in tension will shift the transition between the two limits: *i.e.*, with increased tension, the plate (disk) limit (where anisotropy can be exemplified) will be reduced with the transition regime (whose location can be approximated by the intercept of the two limits) receding towards smaller diameters (Fig. 2c) and larger thicknesses (Fig. 2d). This suggests that in terms of a device's resonant response, the mechanical anisotropy is more accessible when the device tension is low. Therefore, in order to observe and exploit mechanical anisotropy in the frequency domain, device fabrication techniques that produce less tension in the 2D layer are desirable.

Among black P's mechanical properties, besides mechanical anisotropy, one key attribute is its low elastic modulus (compared with other 2D materials, such as graphene), making it much easier to generate high strain in the layered material [7,8,9]. Here, we compare black P nanomechanical resonators with devices made of graphene, the hallmark 2D material. Figure 2e & f shows the frequency scaling of graphene drumhead resonators, analogous to Fig. c & d (same device dimensions and same 2D tension levels). By comparing the two sets of plots, one can see that black P devices are much more readily to operate in the membrane limit—*e.g.*, the transition takes place at smaller $d$ (Fig. 2c *vs.* 2e) and larger $t$ (Fig. 2d *vs.* 2f). It again shows that in order to observe mechanical anisotropy through mechanical resonance, it is important to properly design the device geometry so the black P resonator can operate in the disk limit, which may take a stronger requirement than devices made from other 2D materials.

While mechanical anisotropy plays an important role in the frequency scaling, it does not affect the mode shape of the fundamental resonance mode for circular drumhead devices. In contrast, in higher resonance modes, the effects of mechanical anisotropy are more readily observed. Figure 3 plots the frequency and mode shape of the first 10 resonance modes, obtained through FEM (using COMSOL, see Methods for details), in a family of circular drumhead resonators with the same dimension ($d$=10μm, $t$=200nm), which is in the plate (disk) limit (Fig. 2b). The results are shown for devices made of black P (black curve), and three hypothetical isotropic materials with elastic properties equivalent to that of black P's stiff axis (green curve), soft axis (red curve), and 'averaged' of the two axis (blue dashed curve). Several major differences can be observed between black P and isotropic devices. First, mechanical anisotropy removes the degeneracy in multimode resonance of circular drumhead devices. The otherwise degenerate mode pairs in isotropic devices (*e.g.*, modes 2 & 3, 4 & 5, *etc.*) no longer exist in black P resonators, as the frequencies of the two modes within each pair deviate from each other. With the symmetry breaking due to mechanical anisotropy, the device geometry alone, though still symmetric, can no longer warrant degeneracy between the resonance modes. Second, unlike in isotropic circular devices where the mode shapes take random orientations (at least mathematically, as all directions are equivalent in circular devices), in black P resonators the mode shapes take clear orientation preference, reflecting the underlying intrinsic anisotropy. For example, in mode 2 the two motional domains (individual singly-connected continuous areas enclosed by nodal lines) are aligned in the direction of the soft axis, which in mode 3 the orientation rotates by 90°. This causes the shorter dimension of the motional domains (which plays a more important role in determining the frequency) to align with the stiff axis, and thus mode 3 has higher frequency than mode 2. Similar contrast can be observed between mode 4 and mode 7, both of which has 3 motional domains in a row, with greater frequency difference. Third, new mode shapes (unavailable in isotropic devices) emerge in black P resonators (*e.g.*, mode 4 & 7 in black P resonator, with 3 motional domains in a row), while some isotropic mode shapes disappear (*e.g.*, mode 6 in isotropic devices, the "donut" mode). Other mode shapes deform and exhibit clear orientational preference in domain arrangement and shape (*e.g.*, mode 8 & 10 in black P resonator *vs.* mode 7 & 8 in isotropic devices) as a manifestation of the mechanical anisotropy. The anisotropic modes shapes are analogous to those found in resonators based on isotropic materials with elliptic device geometry.

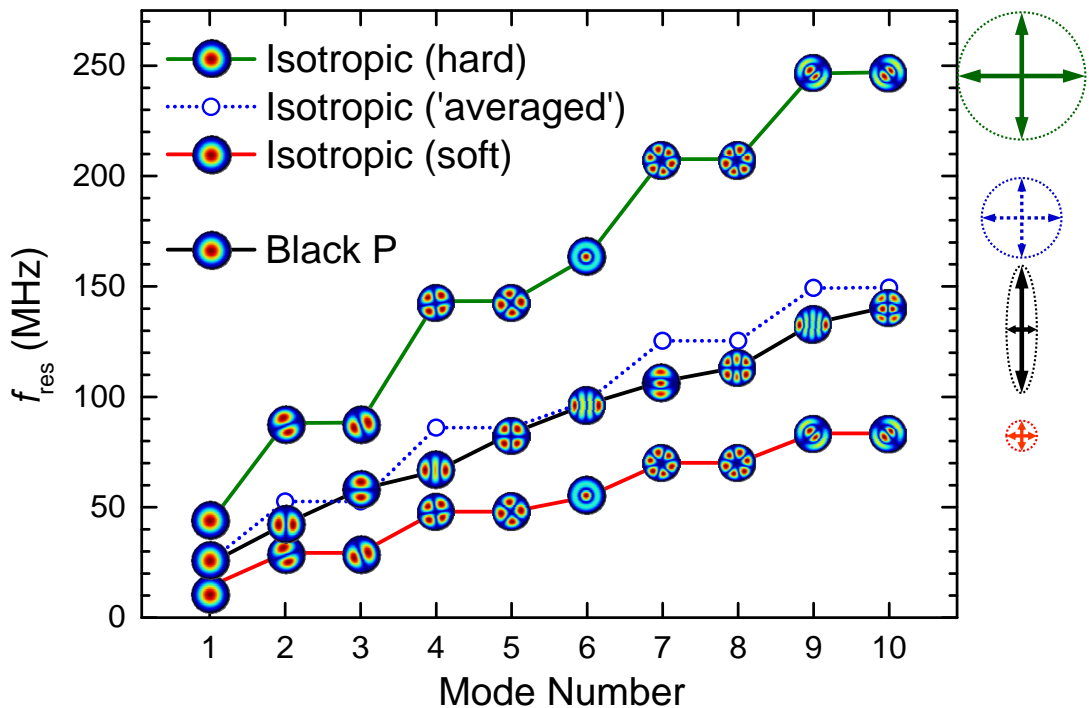

**Figure 3. Frequency scaling of higher mode resonances in circular black P nanomechanical resonators.** The $f_{res}$ and mode shapes for the first 10 modes are shown for circular drumhead resonators with $d$=10μm and $t$=200nm made of black P (black), and hypothetical 2D nanomaterial with its elastic modulus equal to that of black P's soft axis (red), stiff axis (green), and 'averaged' between the two axes (blue). The elastic moduli of these 4 different materials are illustrated to the right, with the arrows represent the magnitude of the moduli in the two in-plane directions (the oval/circles are guides to the eye). The mode shapes for the isotropic device in the middle (blue curve) are similar to those of the soft (red) and stiff (blue) materials, and are thus not shown.

We now turn to the more complicated case of anisotropic device geometry. The basic concept of a nanomechanical resonator with both anisotropic elastic properties and asymmetric device geometry is illustrated in Fig. 4: when the device's orientation (in contrast to the previous circular cases, here a well-defined device orientation exists) is rotated with respected to the crystal axis, the governing equation of motion for the resulting devices obey different boundary conditions, and thus different mechanical responses ensue.

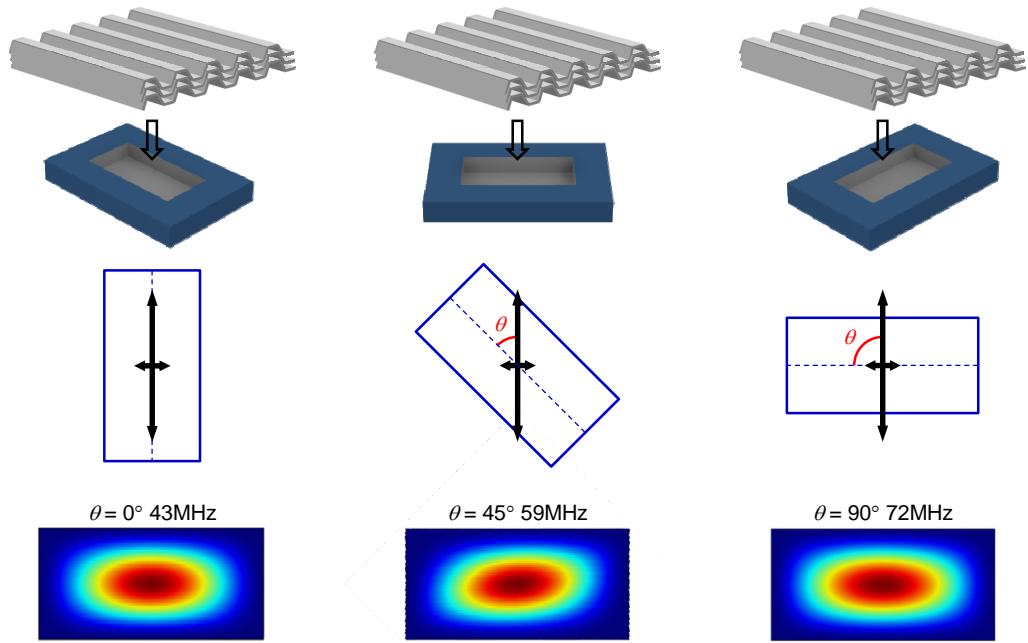

**Figure 4. Effect of crystalline orientation on rectangular black P resonators.** The schematic shows the different relative orientation between the crystal and the device shape, resulting in different alignments between the black P's mechanical stiff axis with the rectangles long side. The FEM simulated mode shapes and $f_{res}$ are shown for three orientations. The arrows represent the magnitude of the elastic moduli in the two in-plane directions (the ovals are guides to the eye). The orientation angle $\theta$ is defined in the illustration.

Here, we examine rectangular drumhead resonators with 2:1 lateral aspect ratio, and use this geometry as an example to demonstrate the effects of mechanical anisotropy in 2D nanomechanical resonators. There are several advantages of choosing a 2:1 rectangle in this study. First, its symmetry belongs to the $D_2$ point group, same as the elasticity modulus profile of black P [7], by having reflection symmetry along two orthogonal axes, and 2-fold (180°) in-plane rotational symmetry about its center. It has the simplest symmetry in 2D polygons, and the relative orientation of the black P crystal to the device geometry can be described by one simple parameter, $\theta$, the angle between the crystalline "stiff" axis and the rectangle's long axis (Fig. 4). Second, it doesn't generate additional symmetry axes as the other polygons (such as triangle or pentagon) do, which may complicate or even obscure the intrinsic effect from the material's mechanical anisotropy. Third, by choosing the 2:1 aspect ratio, the device geometry is sufficiently distinct from a square (which would also introduce additional symmetries and complicate the results), while maintaining a moderate aspect ratio such that the different resonant modes are sufficiently close to each other, allowing us to investigate multimode response (such as mode crossing and mode shape evolution) in black P nanomechanical resonators.

As we demonstrated in the case of circular drumhead resonators, the effect of mechanical anisotropy is only manifested in the nanomechanical resonances when the device operates in the plate (disk) limit, in which the elastic properties dominate the resonant response. Accordingly, we directly investigate the rectangular resonators with dimensions that ensure plate-like behavior. Similar to the circular case, expressing the resonance frequencies of rectangular anisotropic plates in explicit analytical expressions is not practical, even without considering the fact that the relative orientation between the crystal axes and the device shape needs to be computed at every angle. Therefore, we use FEM to study the multimode resonant behavior of rectangular black P resonators. We choose a device geometry of 5μm×10μm with 200nm thick black P, which is in the plate regime (Fig. 2b).

In contrast to circular black P devices, in which the fundamental resonance mode exhibits no sign of anisotropy, in rectangular black P resonators even the first mode show clear anisotropic effects (Fig. 5a): both $f_{res}$ value and the mode shape change as $\theta$ varies, and only at $\theta = n \times 90°$ ($n$=0, 1, 2…) does the mode shape coincide that of an isotropic rectangular device. This strong orientational dependence becomes more evident when we examine the multimode behavior. Figure 5 summarizes the mode shapes and $f_{res}$ values for the first 6 resonance modes as $\theta$, the relative orientation between the crystalline stiff axis and the rectangular long side, varies between 0° and 180°. From the evolution behavior of each mode we make the following observations. First, the resonant modes evolve differently as $\theta$ varies. For example, at 90° some mode has highest $f_{res}$ (*e.g.*, the lowest mode, blue curve), while others may be at the lowest $f_{res}$ (*e.g.*, the mode with 4 motional domains in a row, dark cyan curve at 90°). Some mode even has non-monotonic $f_{res}$ versus $\theta$ relation between 0° and 90°(*e.g.*, the mode with 3 motional domains in a row, dark yellow curve). Second, the mode sequence changes as $\theta$ varies. We find that the mode with only 1 active domain (blue curve) remains the lowest mode, as expected. The mode with two active domains aligned along the long direction of the rectangle (red curve) also remains the 2$^{nd}$ mode throughout. However, the 3$^{rd}$ lowest mode at 0°, with two active domains aligned along the rectangle's short axis, undergoes a significant $f_{res}$ increase as $\theta$ increases, and peaks at 90° where it becomes the 6$^{th}$ mode, with 3 other modes getting below it. Third, as $f_{res}$ of the different modes cross each other, two types of behavior are observed. When the individual mode shapes are retained, the two modes simply cross each other (Fig. 5b). In contrast, if the mode shapes of the two approaching modes gradually deforms and evolve towards each other, the $f_{res}$ of the two modes will avoid crossing (anti-crossing) (Fig. 5c), while the two modes exchange their mode shapes and evolve into each other without becoming degenerate (in $f_{res}$). We find that only certain modes can exhibit anti-crossing behavior: for example, a mode with 2 active domains cannot evolve into one with 3 or 5 domains.

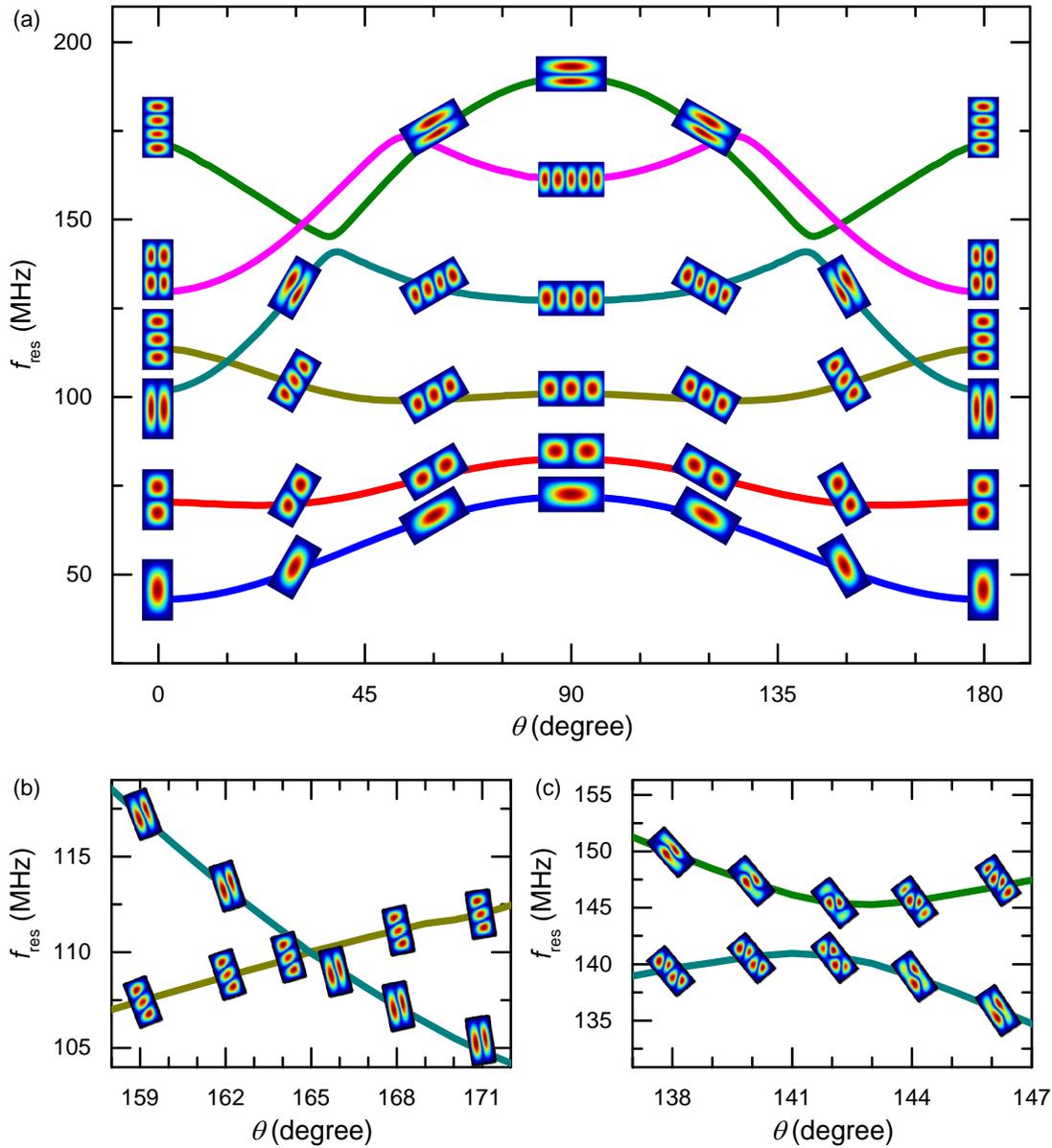

**Figure 5. Evolution of resonant response as a function of crystal orientation in rectangular black P resonators.** (a) The $f_{res}$ and mode shapes for the first 6 modes are shown for 5μm×10μm rectangular black P resonators with $t$=200nm as functions of the orientation angle $\theta$. (b) The zoom-in view of a simple crossing between two modes. (c) The zoom-in view of an anti-crossing between two modes

To explore and understand the role of anisotropy plays in this multimode evolution as the device shape rotates with respect to the material's crystalline axis, we study frequency scaling of the 0° and 90° resonance modes as a function of material's mechanical anisotropy. Specifically, we study nanomechanical resonators based on a family of hypothetical layered materials with Young's modulus ($E_Y$) in the $x$ direction equal to that along black P's soft axis, and with varying $E_Y$ in the $y$ direction. Figure 6a shows how the 0° (solid line) modes and 90° (dashed line) modes evolve as the y direction $E_Y$ increases. Frequency of all modes increases—as the overall elastic modulus of the material increases (one axis fixed, the other increasing)—but at different rates. For some mode shapes, the 0° mode is more sensitive to the $E_Y$ change, while vice versa for the others. For all mode shapes, the 0° and 90° resonance modes degenerate when the material becomes isotropic (same $E_Y$ in both directions, vertical dashed line), as expected. From Fig. 6a one can see that black P is a highly anisotropic material (far away from the isotropic position on the horizontal axis), and thus has mode sequence very different from that of an isotropic material. To disentangle the effect from the material's overall stiffening (Fig. 6a) and focus on the anisotropic effect, in Fig. 6b we plot the $f_{res}$ ratio (instead of values) between the 90° and 0° resonance modes ($f_{90°}/f_{0°}$) as a function of the elastic modulus ratio between the two axes ($E_{Yy}/E_{Yx}$). For clarity we show the results between $0.5 < E_{Yy}/E_{Yx} < 1$ (from data in the shaded region of Fig. 6a). All the mode ratios equal to 1 at $E_{Yy}/E_{Yx} = 1$ (isotropic material). As the mechanical anisotropy increases, the $f_{res}$ ratio varies, with different magnitudes and polarities for the different mode shapes.

Such behavior can be understood by examining the arrangement and shape of individual motional domains inside each mode shape. For example, in the fundamental mode (black curve), the motional domain is close to the shape of the entire device (2:1 rectangle). When $E_{Yy} < E_{Yx}$, the stiff axis is aligned with the long side of the rectangle in the 90° mode, and with the short side in the 0° mode. Accordingly, $f_{90°} < f_{0°}$ when $E_{Yy}/E_{Yx} < 1$, and vice versa. The second lowest mode (red curve) has two active domains along the long axis. Consequently, the shape of each domain is close to a square, and is thus less sensitive to the change in $E_{Yy}/E_{Yx}$ (as the two axes become more or less equivalent). In contrast, the mode with two active domains arranged along the short axis (magenta curve) exhibit very strong dependence, as the shape of each motional domain is highly asymmetric (close to 4:1 rectangle) and is thus highly sensitive to the crystalline orientation of the anisotropic material. The modes with three (blue curve) and four (dark yellow curve) motional domains along the rectangular long axis exhibit opposite anisotropic dependence as the others, because the orientation of each of their motional domains is exactly orthogonal to that of the entire device (*i.e.*, the long axis direction). Accordingly, their frequency ratio dependence on the crystal orientation becomes the opposite as other modes.

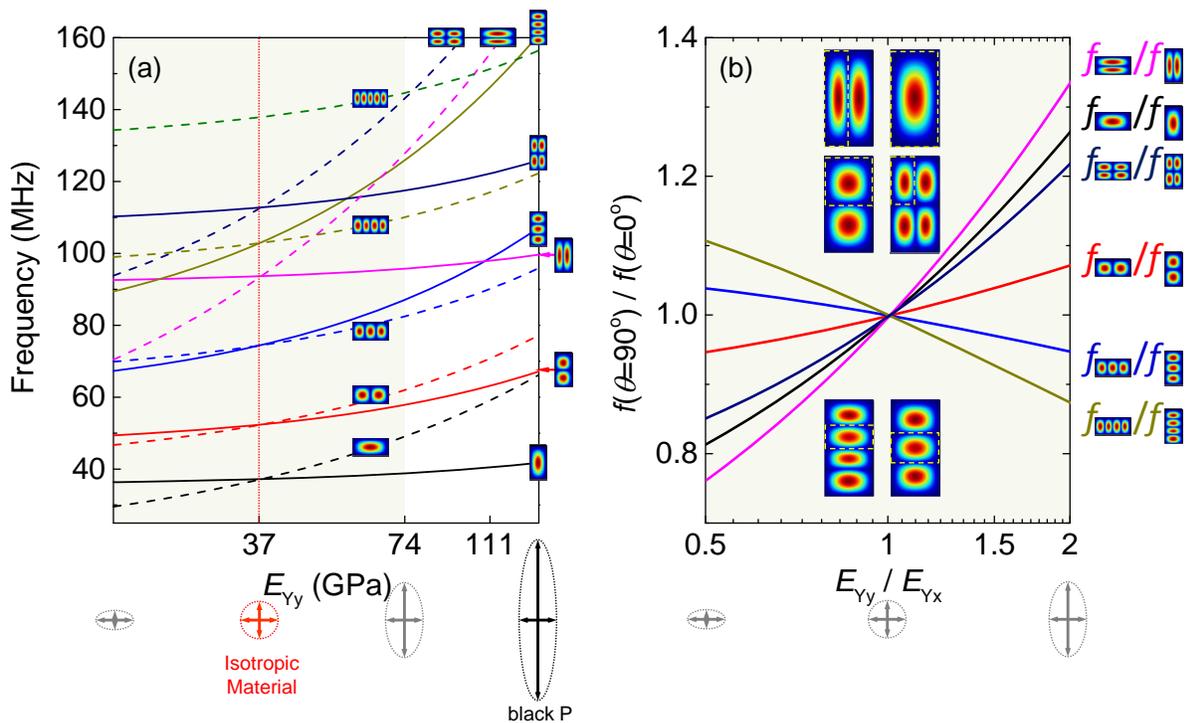

**Figure 6. Evolution of multimode resonant response with degree of mechanical anisotropy.** (a) Resonance frequency as a function of the Young's modulus in the y direction for a 10μm×5μm×200nm black P nanomechanical resonator. Solid lines show $f_{res}$ for $\theta$=0° (the rectangle's long side along y direction), and dashed lines show results for $\theta$=90°. The elastic parameters for black P and isotropic material are indicated on the horizontal axis, with the crossing arrows representing the magnitude of the elastic moduli in the two in-plane directions (the ovals are guides to the eye). (b) $f_{res}$ ratio (between the 90° and 0° arrangements) as a function of the ratio of Young's moduli in the two directions. The data range corresponds to the shaded area in (a). *Insets*: individual mode shapes with corresponding motional domain highlighted (dashed boxes).

In summary, the resonant response in black P nanomechanical resonators has been theoretically investigated. For circular drumhead devices, different elastic regimes are delineated through studying the frequency scaling of the fundamental mode by varying device thickness, diameter, and 2D tension. The results are used to identify the parameter space where anisotropy may be observed through mechanical resonances. The effect of mechanical isotropy is more readily observed in the higher resonance modes through mode shape as well as mode sequence. For rectangular resonators, the effect of device orientation on the multimode frequency scaling, and the dependence on the degree of mechanical anisotropy for the different resonant modes are explored. The results show that black P and other anisotropic layered materials open opportunities for designing and building novel nanomechanical devices which can offer anisotropic multimode resonant responses unavailable in devices based on isotropic materials.

**Methods**

The parameters for FEM simulation are chosen from values reported in the literature.

The in-plane elastic moduli are 159GPa and 37GPa, following the results for 4L black P in Ref. 7 (the values vary little with the number of the layers, and here we treat them as constants in multilayer black P). The out-of-plane elastic modulus is chosen to be 40GPa, an average number of the black P $C_{33}$ bulk modulus values calculated using different potentials [7,9]. The FEM results are not sensitive to the choice of this value due to the high aspect ratio in the device structures.

The Poisson's ratio for the in-plane directions are 0.73 and 0.17, following Ref. 7. For the out-of-plane direction we use a Poisson's ratio of 0.3, following Ref. 9. We notice that different values of Poisson's ratio are reported in the literature, including negative ones [30]. We found that the FEM result is not very sensitive to the choice of Poisson's ratio, and does not qualitatively change even when the frequencies vary. The shear modulus is chosen to be 45GPa [7] for all three directions.

The hypothetical 'averaged' isotropic material used to compare with black P (in Fig. 3, blue curve) has Young's modulus of 100GPa and Poisson's ratio of 0.4. The values are determined such that the resonance frequency for the fundamental mode best matches that of black P circular drumhead resonators in the entire $d$=1−10μm and $t$=10−500nm range.

**Acknowledgement:** We thank X.-Q. Zheng for help with the illustrations. We thank the support from Case School of Engineering, National Academy of Engineering (NAE) Grainger Foundation Frontier of Engineering (FOE) Award (FOE2013-005), and National Science Foundation CAREER Award (ECCS #1454570).